\documentclass[onecollarge,natbib]{svjour2}
\bibpunct{[}{]}{;}{n}{}{,} 
\smartqed  
\usepackage{graphicx}
\journalname{Few-Body Systems (EFB22)}
\begin{document}

\title{ Phenomenological studies of the low energy dynamics in the $ppK^+K^-$ system
}


\author{M. Silarski
}


\institute{M. Silarski \at
              Institute of Physics, Jagiellonian University, 30-059 Krak{\'o}w, Poland \\
              Tel.: +48126635619\\
              Fax: +48126637086 \\
              \email{michal.silarski@uj.edu.pl}           
}

\date{Received: date / Accepted: date}

\maketitle

\begin{abstract}
In this article we review studies of the near threshold $pp\to ppK^+K^-$ reaction
done with the COSY-11 and ANKE detectors. We discuss phenomenological studies of the $ppK^+K^-$ dynamics,
in particular the recent investigations on the $K^+K^-$ final state interaction are presented.
\keywords{final state interaction \and near threshold kaon pair production}
\end{abstract}
\section{Introduction}
\label{intro}
The primary motivation for the near threshold $pp\to ppK^+K^-$ reaction studies was closely connected
with understanding of the nature of scalar resonances $a_0(980)$ and $f_0(980)$, which properties suggest
that they are not ordinary mesons. Besides the interpretation as $q\bar{q}$ mesons~\cite{Morgan}, these
particles were also proposed to be $qq\bar{q}\bar{q}$ tetraquark states~\cite{Jaffe}, hybrid $q\bar{q}$/meson-meson
systems~\cite{Beveren} or even quark-less gluonic hadrons ~\cite{Johnson}. Since both $f_{0}$(980) and $a_{0}$(980)
masses are very close to the sum of the $K^{+}$ and $K^{-}$ masses, they are considered also as $K\bar{K}$ bound
states~\cite{Lohse, Weinstein}, which formation requires a strong $K\bar{K}$ interaction.
Since kaon targets are still unavailable, the $K^+K^-$ potential can be probed experimentally only in a multi-particle
production reactions, like e.g. the $pp\to ppK^+K^-$ near threshold, where in the first order the S-wave interaction
is relevant. Studies of the low energy $pp\to ppK^+K^-$ system gives opportunity to investigate also the $pK^-$ final state
interaction, very important in view of the nature of $\Lambda(1405)$ which is often considered as the $N$-$K^-$ molecule.\\
Measurements of the near threshold $pp\to ppK^+K^-$ reaction have been performed mainly at the cooler synchrotron
COSY at the research center in J{\"u}lich, Germany~\cite{cosy}. COSY provides proton and deuteron beams with low
emittance and small momentum spread. This allowed for precise determination of the proton-proton collision energy,
in the order of fractions of MeV, and for dealing with the rapid growth of cross section at the threshold.
First measurements of the $pp\to ppK^+K^-$ reaction were performed by the COSY-11 collaboration and
they revealed that the total cross sections near threshold are relatively small (in the order of nanobarns)
making the study difficult due to low statistics~\cite{wolke,quentmeier,winter}. Moreover, a possible influence
from the $f_{0}$ or $a_{0}$ mesons on the $K^{+}K^{-}$ pair production appeared to be too weak to be distinguished
from the direct production of these mesons based on the COSY-11 data~\cite{quentmeier}.
However, the ratios of the differential cross sections as a function of the $pK$ and
the $ppK$ invariant masses measured at excess energies $Q=$~10~MeV and $Q=$~28~MeV:
\begin{eqnarray}
\nonumber
R_{pK} &= \frac{\mathrm{d}\sigma/\mathrm{d}M_{pK^{-}}}{\mathrm{d}\sigma/\mathrm{d}M_{pK^{+}}}~,
\nonumber
R_{ppK} &= \frac{\mathrm{d}\sigma/\mathrm{d}M_{ppK^{-}}}{\mathrm{d}\sigma/\mathrm{d}M_{ppK^{+}}}~,
\end{eqnarray}
showed a significant enhancement in the region of both the low $pK^-$ invariant mass $M_{pK^{-}}$
and the low $ppK^-$ invariant mass $M_{ppK^{-}}$~\cite{winter,PhysRevC}. Since the $pK^+$ interaction
is known to be very weak, this enhancement indicates a strong influence of the $pK^-$ final state interaction.
This effect was confirmed by the ANKE collaboration at significantly higher energies~\cite{anke,Ye,anke_last}.
Influence of the interaction in the low energy $ppK^+K^-$ system manifests also in the shape of the $pp \to ppK^+K^-$
excitation function presented in Fig.~\ref{funkcja_wzbudzenia}, where one observes a strong deviation from the pure
phase space expectations (dashed curve). 
\section{Description of the dynamics in the low energy $ppK^+K^-$ system}
\label{sec:2}
\begin{figure}
\centering
\includegraphics[width=0.47\textwidth]{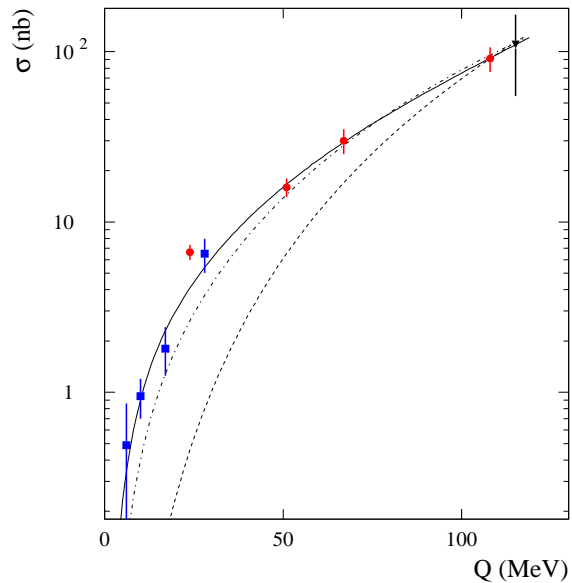}
\caption{Excitation function for the $pp\rightarrow ppK^{+}K^{-}$ reaction.
Triangle and circles represent the DISTO and ANKE measurements, respectively~\cite{anke,anke_last,disto}.
The squares are results of the COSY--11~\cite{wolke,quentmeier,PhysRevC} measurements.
The dashed curve represents the energy dependence
obtained assuming that the phase space is homogeneously and isotropically populated,
and there is no interaction between particles in the final state.
Calculations taking into account the $pp$ and
$pK^-$ FSIs are presented as the dashed--dotted curve. 
The dashed and dashed--dotted curves are
normalized to the DISTO data point at Q~=~114~MeV. 
Solid curve corresponds to the result obtained taking into account
$pp$, $pK$, and $K^+K^-$ interactions parametrized with the effective range approximation.
These calculations were obtained using the quoted scattering length $a_{K^{+}K^{-}}$ and effective
range $b_{K^+K^-}$ values. Figure adapted from~\cite{physRevC2}.}
\label{funkcja_wzbudzenia}
\end{figure}

In the close-to-threshold region kaon pairs production requires large momentum transfer between the interacting nucleons,
thus the complete transition matrix element of the $pp\to ppK^+K^-$ reaction may be factorized approximately as~\cite{moskal}:
\begin{equation}
\left|M_{pp\to ppK^+K^-}\right|^2 \approx \left| M_{0}\right|^2\cdot \left| M_{FSI}\right|^2~,
\end{equation}
where $\left| M_{0}\right|^2$ represents the total short range production amplitude, and $\left| M_{FSI}\right|^2$
denotes the final state interaction enhancement factor. Since exact calculations for the dynamics of four-body
final states are still unavailable, the enhancement factor for the $ppK^+K^-$ system was approximated
assuming the factorization of $M_{FSI}$ to the two-particle scattering amplitudes~\cite{wilkin}:
\begin{equation}
M_{FSI} = F_{pp}(k_{1}) \times F_{p_{1}K^-}(k_{2}) \times F_{p_{2}K^-}(k_{3})~,
\label{row1}
\end{equation}
where $k_{1}$, $k_{2}$ and $k_{3}$ denote the relative momentum of particles in the proton-proton and
two proton-$K^-$ subsystems.
Here one assumes strong proton-proton and $pK^-$ interactions neglecting the $K^+$
influence. Using this approximation the ANKE collaboration have described all the differential distributions
measured at higher energies from the threshold using an effective scattering length $a_{pK^-} = i$1.5~fm~\cite{anke}.
This model, however, underestimates COSY-11 total cross sections near threshold, as one can see in Fig.~\ref{funkcja_wzbudzenia} 
(dashed--dotted curve), which indicates that in the low energy region the influence of the $K^{+}K^{-}$
final state interaction may be significant.\\
Motivated by this observation the COSY-11 collaboration has estimated the scattering length of the $K^{+}K^{-}$
interaction  based for the first time on the low energy $pp \to ppK^{+}K^{-}$ Goldhaber Plot
distributions measured at excess energies of Q~=~10~MeV and 28~MeV~\cite{PhysRevC}.
The final state interaction model used in that analysis is was based on the factorization
ansatz in Eq.~\ref{row1}, with an additional term describing the interaction of the $K^+K^-$
pair.
Factors describing the enhancement originating from the $pK^-$
and $K^+K^-$--FSI were instead parametrized using the scattering length approximation:
\begin{equation}
\nonumber
F_{pK^{-}}=\frac{1}{1-ika_{pK^-}}~,~~~F_{K^{+}K^{-}}=\frac{1}{1-ik_{4}~a_{K^+K^-}}~,
\label{F_ppKK}
\end{equation}
where $a_{pK^-} = i$1.5 fm  and $a_{K^+K^-}$ is the scattering length of the $K^{+}K^{-}$
interaction treated as a free parameter in the analysis. 
As a result of these studies $a_{K^+K^-}$ was estimated to be:
$\left|Re(a_{K^{+}K^{-}})\right| = 0.5^{~+4}_{~-0.5}$~fm and $Im(a_{K^{+}K^{-}}) = 3~\pm~3$~fm.\\
Within this simple model any coupled channel effects, like e.g. the charge-exchange
interaction allowing for the $K^{0}K^{0}\rightleftharpoons K^{+}K^{-}$
transitions, were neglected. However, studies done by the ANKE collaboration showed that even with
their high statistics data can be described well without introducing these effect~\cite{dzyuba}.
\section{Determination of the effective range of the $K^+K^-$ final state interaction}
\label{sec3}
Since the $pK^{-}$ scattering length estimated by the ANKE group is rather an effective
parameter~\cite{anke}, we have repeated the $K^+K^-$ final state interaction studies
using more realistic $a_{pK^-}$ value estimated independently as a mean of all values
summarized in Ref.~\cite{Yan:2009mr}: $a_{pK^-} = (-0.65 + 0.78i$)~fm. Moreover, in the fit
we have taken into account not only the differential cross sections measured by the COSY-11 collaboration,
but also all the $pp\to ppK^+K^-$ total cross sections measured near threshold~\cite{eqcd13,physRevC2}. Since the energy
range for the experimental excitation function is rather big the $K^+K^-$ final state enhancement
factor was approximated using the effective range expansion:
\begin{equation}
\nonumber
F_{K^{+}K^{-}}=\frac{1}{\frac{1}{a_{K^+K^-}}+\frac{b_{K^+K^-}k^2_{4}}{2}-ik_{4}},
\label{F_KKb}
\end{equation}
where $a_{K^+K^-}$ and $b_{K^+K^-}$ are the scattering length and the effective range
of the $K^{+}K^{-}$ interaction, respectively.
Moreover, we have repeated the analysis for every $a_{pK^-}$ value quoted  in Ref.~\cite{Yan:2009mr}
to check how their different values change the result. This allowed us also
to estimate the systematic error due to the used $pK^{-}$ scattering length used in the estimation
of $a_{K^+K^-}$ and $b_{K^+K^-}$. The best fit to the experimental data corresponds to
\begin{eqnarray}
\nonumber
\mathrm{Re}(b_{K^{+}K^{-}}) = -0.1 \pm 0.4_{stat} \pm~0.3_{sys}~\mathrm{fm}\\
\nonumber
\mathrm{Im}(b_{K^{+}K^{-}}) = 1.2^{~+0.1_{stat}~+0.2_{sys}}_{~-0.2_{stat}~-0.0_{sys}}~\mathrm{fm},\\
\nonumber
\left|\mathrm{Re}(a_{K^{+}K^{-}})\right| = 8.0^{~+6.0_{stat}}_{~-4.0_{stat}}~\mathrm{fm}\\
\mathrm{Im}(a_{K^{+}K^{-}}) = 0.0^{~+20.0_{stat}}_{~-5.0_{stat}}~\mathrm{fm},
\label{chi2resultsB}
\end{eqnarray}
with a $\chi^2$ per degree of freedom of: $\chi^2/ndof = 1.30$~\cite{physRevC2}.
The fit is in principle sensitive to both the scattering length and effective range,
however, with the available low statistics data the sensitivity to the scattering length is very weak.
Calculations with inclusion of the interaction in the $K^+K^-$ system
described in this section are shown as the solid curve in Fig.~\ref{funkcja_wzbudzenia}.
One can see that the experimental data are described quite well over the whole energy range. 
\section{Summary and outlook}
Studies of the $pp\to ppK^+K^-$ reaction near threshold done with the COSY-11 and ANKE detectors
suggest that the scalar resonances contribution to this reaction is negligible. The enhancement
in the total cross sections may be explained introducing final state interaction, which manifests
itself for example in the ratios of differential cross sections as a function of the $pK$ and
the $ppK$ invariant masses. The influence of the $pK^-$ interaction is seen in the broad range
from very close to threshold energy up to Q = 108 MeV~\cite{PhysRevC,anke}. Dynamics in the $ppK^+K^-$
system produced near threshold was described within a simple model assuming factorization of the total
final state interaction enhancement factor into two-particles interactions. Within this model we have
estimated parameters of the $K^+K^-$ interaction, in particular the effective range of this interaction.\\
However, the latest ANKE results obtained at $Q$~=~24 MeV suggest that the interaction in the $ppK^+K^-$
system is far more complex and one should use much more sophisticated model than the factorization
anszatz~\cite{anke_last}. Thus, the results of analysis quoted in this article should be considered
rather as effective parameters. It seems that this reaction is driven by the $\Lambda(1405)$ production
$pp \to K^+\Lambda(1405) \to ppK^+K^-$ rather than by the scalar mesons~\cite{dzyuba} with possible cusp
effect generated by the $pK^- \rightleftharpoons n\overline{K}^0$ coupling.
\begin{acknowledgements}
The author is grateful to P.~Moskal and E.~Czerwi{\'n}ski for their valuable comments and corrections.
This research was supported by the FFE grants of the Research Center J{\"u}lich,  
by the Polish National Science Center and by the Foundation for Polish Science.
\end{acknowledgements}


\begin{thebibliography}{3}
%
\bibitem[Morgan (1993)]{Morgan}
Morgan,~D., Pennington, ~M.R.:New data on the $K\overline{K}$ threshold region and the nature of the $f_0(S*)$.
Phys. Rev.~D 48, 1185-1204 (1993).
\bibitem[ Jaffe (1977)]{Jaffe}
Jaffe,~R.~L.: Multiquark hadrons. I. Phenomenology of $Q^2\overline{Q}^2$ mesons.~Phys. Rev.~D 15, 267-280 (1977).
\bibitem[Beveren (1986)]{Beveren}
Van Beveren,~E.~\textit{et al.}: A low lying scalar meson nonet in a unitarized meson model. Z. Phys.~C~30, 615-620 (1986).
\bibitem[Jaffe (1976)]{Johnson}
Jaffe, R. L., Johnson, K.: Unconventional states of confined quarks and gluons. Phys. Lett. B 60, 201-204 (1976).
\bibitem[Lohse (1990)]{Lohse}
Lohse, D.~\textit{et al.}: Meson exchange model for pseudoscalar meson-meson scattering. Nucl. Phys. A 516, 513-548 (1990).
\bibitem[Weinstein (1990)]{Weinstein}
Weinstein, J. D., Isgur, N.: $K\overline{K}$ molecules.~Phys. Rev.~D 41, 2236-2257 (1990).
\bibitem[Prasuhn (2000)]{cosy}
Prasuhn, D.~\textit{et al.}: Electron and stochastic cooling at COSY. Nucl. Instrum. Methods Phys. Res.~A 441, 167-174 (2000).
\bibitem[wolke (1997)]{wolke}
Wolke, M.: Schwellennahe assoziierte Strangeness-Erzeugung in der Reaktion $pp\to ppK^+K^-$ am Experiment COSY-11. IKP J{\"u}l-3532 (1997).
\bibitem[quentmeier (2001)]{quentmeier}
Quentmeier, C.~\textit{et al.}: Near threshold $K^+K^-$ meson-pair production in proton–proton collisions. Phys. Lett. B 515, 276-282 (2001).
\bibitem[winter (2006)]{winter}
Winter, P.~\textit{et al.}: Kaon pair production close to threshold. Phys. Lett. B 635, 23-29 (2006).
\bibitem[silarski (2009)]{PhysRevC}
Silarski, M.~\textit{et al.}, Generalized Dalitz Plot analysis of the near threshold $pp\rightarrow ppK^{+}K^{-}$
reaction in view of the $K^{+}K^{-}$ final state interaction. Phys. Rev.~C 80, 045202 (2009).
\bibitem[maeda (2008)]{anke}
Maeda, Y. {\it et al.}: Kaon Pair Production in Proton–Proton Collisions. Phys. Rev. C 77, 015204 (2008).
\bibitem[Ye (2012)]{Ye}
Ye, Q.~J.~\textit{et al.}: The production of $K^+K^-$ pairs in proton-proton collisions at 2.83 GeV.
Phys. Rev. C 85, 035211 (2012).
\bibitem[Ye (2013)]{anke_last}
Ye, Q.~J. {\it et al.}: The production of $K^+K^-$ pairs in proton-proton collisions below the $\phi$ meson threshold.
Phys. Rev. C 87, 065203 (2013).
\bibitem[moskal (2004)]{moskal}
Moskal, P.: Hadronic interaction of $\eta$ and $\eta'$ mesons with protons. Jagiellonian University Press, Krak{\'o}w (2004).
\bibitem{wilkin}
Wilkin, C.: Coupled-Channel Effects in Kaon Pair Production, Acta Phys. Polon. B Proceed. Supp. 2, 89-96 (2009).
\bibitem[dzyuba (2008)]{dzyuba}
Dzyuba, A.~\textit{et al.}: Coupled-channel effects in the $pp\to ppK^+K^-$ reaction.
Phys. Lett. B 668, 315-324 (2008).
\bibitem[yan (2009)]{Yan:2009mr}
Yan, Y.: Kaonic hydrogen atom and $K^-p$ scattering length. arXiv:0905.4818 (2009).
\bibitem[silarski (2013)]{eqcd13}
Silarski, M.: Study of the $K^+K^-$ final state interaction in proton-proton and electron-positron collisions.
Acta Phys. Polon. B Proceed. Suppl. 6, 865-871 (2013).
\bibitem[silarski (2013)]{physRevC2}
Silarski, M., Moskal, P.: Combined analysis of the $K^{+}K^{-}$ interaction using near threshold
$pp \to ppK^+K^-$ data. Phys. Rev. C 88, 025205 (2013).
%
\bibitem{disto}
Balestra, F.~\textit{et al.}: $K^-$ meson production in the proton–proton reaction at 3.67 GeV/c.
Phys. Lett. B 468, 7-12 (1999).
%
\end{thebibliography}
\end{document}